\documentclass[useAMS,usenatbib]{mn2e}

\usepackage{lscape}
\usepackage{graphicx}
\usepackage{natbib}
\usepackage{graphics}

\def \vhel{\ifmmode{~V_{{\rm HEL}}}\else{~$V_{{\rm HEL}}$}\fi}
\def \vsys{\ifmmode{~V_{{\rm SYS}}}\else{~$V_{{\rm SYS}}$}\fi}
\def \HA {\ifmmode{{\rm\H}\alpha}\else{${\rm\ H}\alpha$}\fi}
\def \farcs{\hbox{$~\!\!^{\prime\prime}$}} 

\def \msun{\ifmmode{{\rm\ M}_\odot}\else{${\rm\ M}_\odot$}\fi}

\def \myr{\ifmmode{{\rm\ M}_\odot{\rm\ yr}^{-1}}
         \else{${\rm\ M}_\odot$ yr$^{-1}$}\fi}

\def \mdot{\ifmmode{\dot{M}}\else{$\dot{M}$}\fi}
\def \tena#1 #2 {\ifmmode{#1 \times 10^{#2}}\else{$#1 \times 10^{#2}$}\fi}
\def \kms{\ifmmode{~{\rm km\,s}^{-1}}\else{~km s$^{-1}$}\fi}

\title[First e-VLBI observations of~GRS\,1915+105]{First e-VLBI observations of~GRS\,1915+105}
\author[A. Rushton et al.]{A. Rushton,$^1$\thanks{E-mail:
arushton@jb.man.ac.uk (AR)} R. E. Spencer,$^1$ M. Strong,$^1$ R. M. Campbell,$^2$ S. Casey,$^1$\and R. P. Fender,$^{3,4}$ M. A. Garrett,$^2$ J. C. A. Miller-Jones,$^4$ G. G. Pooley,$^5$ C. Reynolds,$^2$\and A. Szomoru,$^2$ V. Tudose$^{4,6}$ and Z. Paragi$^2$\\
\\
$^1$The University of Manchester, Jodrell Bank Observatory, Cheshire SK11 9DA\\
$^2$Joint Institute for VLBI in Europe, Postbus 2, 7990 A A Dwingeloo, The Netherlands\\
$^3$School of Physics and Astronomy, University of Southampton, Highfield, SO17 1BJ Southampton, UK\\
$^4$``Anron Pannekoek'' Astronomical Institute, University of Amsterdam, Kruislaan 403, 1098 SJ Amsterdam, The Netherlands\\
$^5$University of Cambridge, Mullar Radio Astronomy Observatory, J. J. Thomson Avenue, CB3 0HE Cambridge, UK\\
$^6$Astronomical Institute of the Romanian Academy, Cutitul de Argint 5 RO-040557 Bucharest, Romania}
\begin{document}

\date{Accepted 2006 October 25.  Received 2006 October 25; in original form 2006 October 6}

\pagerange{\pageref{firstpage}--\pageref{lastpage}} \pubyear{2006}

\maketitle

\label{firstpage}

\begin{abstract}
We present results from the first successful open call e-VLBI science run, observing the X-ray binary GRS\,1915+105. e-VLBI science allows the rapid production of VLBI radio maps, within hours of an observation rather than weeks, facilitating a decision for follow-up observations. A total of 6 telescopes observing at 5~GHz across the European VLBI Network (EVN) were correlated  in real time at the Joint Institute for VLBI in Europe (JIVE). Constant data rates of 128~Mbps were transferred from each telescope, giving 4 TB of raw sampled data over the 12 hours of the whole experiment. Throughout this, GRS\,1915+105 was observed for a total of $5.5$~hours, producing 2.8~GB of visibilities of correlated data. A weak flare occurred during our observations, and we detected a slightly resolved component of 2.7 $\times$ 1.2~milliarcsecond with a position angle of $140^\circ\pm2^\circ$. The peak brightness was $10.2$~mJy~per~beam, with a total integrated radio flux of $11.1$~mJy.

\end{abstract}


\section{Introduction}

\begin{figure*}
\centering
\mbox{\resizebox{11cm}{!}{\includegraphics{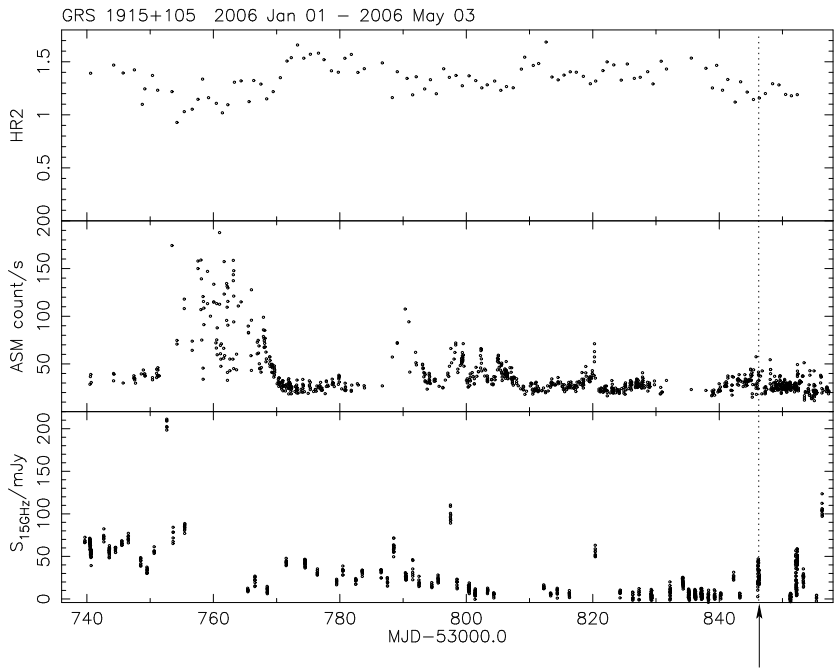}}}
\caption{\label{monitor_grs1915+105}RXTE X-ray and Ryle telescope 15 GHz flux density monitoring of GRS\,1915+105, observed between 2006~Jan~01 and May~03. The data of the e-VLBI observation (MJD 53846), is marked by the dotted line. The top shows the X-ray spectral hardness ratio$\left(\frac{5-12keV}{3-5keV}\right)$, the middle shows the RXTE ASM count-rate and the bottom shows the 15 GHz radio flux density.}
\end{figure*}

The use of the Internet for electronic very-long-baseline interferometry (e-VLBI) data transfer offers a number of advantages over conventional recorded VLBI, including improved reliability due to real time operation and the possibility of a rapid response to new and transient phenomena. Decisions on follow-up observations can be made immediately after the observation rather than delayed by potentially weeks due to problems in shipment of tapes/discs to the correlator. The first open call with a suitable GST range for observations of GRS\,1915+105 using the e-EVN (electronic European VLBI Network)\footnote{http://www.evlbi.org/evlbi} gave us the opportunity to test e-VLBI under operational conditions. A number of recent test runs have shown that 128~Mbps data rates can be obtained reliably to the 6 European telescopes; Cambridge, Jodrell Mk2, Medicina, Onsala, Torun and Westerbork, currently connected via national and international research networks to the EVN correlator at Joint Institute for VLBI in Europe (JIVE). Steps are currently being taken to improve the reliability of 256 and 512~Mbps connections, and also develop 1~Gbps transmission as part of the EXPReS\footnote{\mbox{see -- http://www.expres-eu.org}} project.

Microquasars are ideally suited for study by e-VLBI since they often have flares associated with the ejection of radio emitting clouds in the form of jets. Time-scales of this emission are in the range of hours to days at cm wavelengths, and decisions about subsequent observations, need to be taken quickly.

The X-ray binary GRS\,1915+105 was first discovered in 1992~\citep{1992IAUC.5590....2C} by the WATCH instrument on the GRANAT satellite. The system comprises a low mass, K-M III star \citep{2001A&A...373L..37G} companion and a $14$ $\pm$4 $\msun$ black hole \citep{2001Natur.414..522G}. It was the first Galactic source observed to display superluminal motion, and is well known for its rapid variability and strong variable radio flux. It spends the majority of its time in relative radio-quiescence, with low radio and X-ray brightness, and with a characteristic low/hard state X-ray spectrum. In such a state the source is thought to be `jet-dominated'~\citep{2003MNRAS.343L..99F}, with a $\sim50$~AU scale inner radio jet~\citep{2000ApJ...543..373D} present. Transitions to the soft state are often accompanied by strong radio flares with the ejection of a high velocity component out to distances of several hundred milliarcsecond or $\sim10^4$~AU; these transitions have been studied by the VLA and MERLIN~\citep{1994Natur.371...46M,1999MNRAS.304..865F,2005MNRAS.363..867M}. Long-term high sensitivity VLBI monitoring of motions in the core is necessary to understand how the inner jets relate to the larger scale ejections. This is not possible without the strategy in place enabling rapid decisions on follow-up VLBI observations.

The large scale ejections have apparent superluminal knots or clouds with velocities of $>0.9c$~\citep{2005MNRAS.363..867M}. Observations with MERLIN in March\,/\,April and July 2001 at 5~GHz gave support for the internal shock model~\citep{2000A&A...356..975K}; an increase in the velocity of the jet material forms shocks in the outflow and superluminal knots are observed. Ideally, observing the source during a state change would reveal the most information.

Over the first few months in 2006, GRS\,1915+105 has been consistently flaring in radio (Figure~\ref{monitor_grs1915+105}). A 300 mJy (at 4.8~GHz) steep spectrum, optically thin flare was detected by the RATAN 600 telescope on 2006 Feb 23, suggesting that the source may have undergone a transition to the high\,/\,soft state. This triggered a MERLIN target of opportnity (ToO) on transient sources which detected another outburst in March~2006 (Miller-Jones et al. in prep.). One aim of the project was also to develop a strategy for rapid response (ToO) e-VLBI observations for when this technique is more mature.

\section{Observations and results}

\begin{figure}
\centering
\rotatebox{0}{\mbox{\resizebox{8.6cm}{!}{\includegraphics{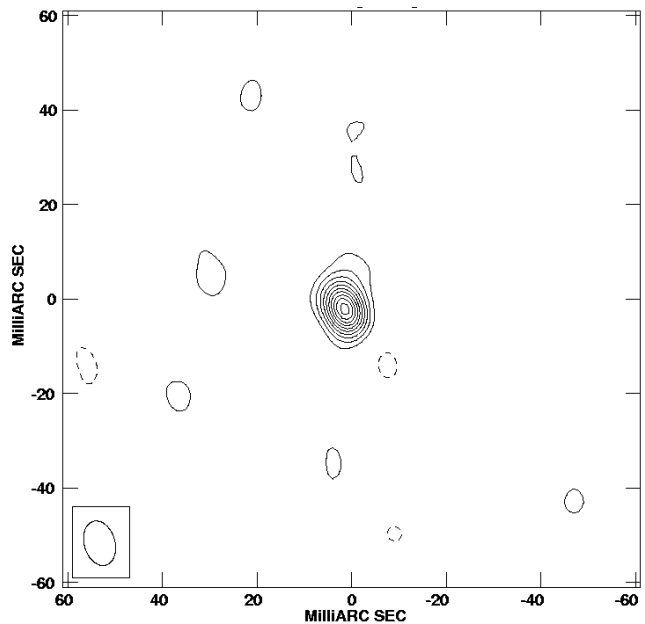}}}}
\caption[grs1915]{\label{grs1915}e-EVN map of GRS\,1915+105 at 5~GHz using 6 telescopes centred at R.A.~$19^{\rm h}\,~15^{\rm m}\,~(11.548\pm0.001)^{\rm s}\,$ and Dec.~$10^{\circ}\,~56^{'}\,~(44.71\pm0.01)\farcs\,$~(J2000) on 21 April 2006. Contour levels are (-1, 1, 2, 4, 6, 8, 10) times 1~mJy per beam, with an rms of 0.3~mJy. The beam size was 9.6$\times$~9.5~mas and was deconvolved with the source revealed an extended component of 2.7 $\times$ 1.2~mas at a position angle of $140~(\pm2)^\circ$.}
\end{figure}

On 2006 April 20~--~21 the e-EVN observed GRS\,1915+105 at 4.994~GHz. This observation was scheduled to be interleaved with a companion project on Cyg X-3 (Tudose et al. submitted) to allow a better \textit{uv}-distribution for both objects in the available observation period. GRS\,1915+105 and its phase-reference source were observed between 23:39 and 10:45 UT for a total time on source of 5.5 hours. 

In this e-VLBI experiment, the data were transferred from the telescope to the correlator using Mark 5A disk-based VLBI data systems. These units have been fitted with 1 Gbps Network Interface Cards which allow the units to transfer the telescope data to the correlator over the Internet and private optical networks at rates exceeding 100~Mbps. Production Internet connections for institutions within each participating country are provided and controlled by the local and national network providers. Most of the telescopes connect to the national networks, and then are connected to the G\'{E}ANT~2 network\footnote{see -- http://www.geant2.net} allowing pan-European multi-gigabit connectivity. The exceptions are Westerbork Synthesis Radio Telescope (WSRT) which has its own direct fibre connection to JIVE,  and Jodrell Bank with a private connection to Manchester followed by a light path to JIVE. SURFnet provids the connection from G\'{E}ANT~2 from Amsterdam to JIVE over multiple optical fibres. Further details on the Internet connections will be presented by Szomoru et al. (in prep.) and Strong et al. (in prep).

Each station sustained a transfer rate of 128~Mbps across the e-VLBI network. This transmission rate supports two 8 MHz dual-polarisation basebands channels, providing a total bandwidth of 32 MHz. The observations were made using the phase-reference mode with a cycle of 5 minutes on source and 3 minutes on the phase reference, J1925+1227. A bright compact radio source, J2002+4725 was used as a fringe finder, was scheduled at the beginning and toward the end of the observing run. 

The initial data reduction was performed using the NRAO software package \textsc{aips}. The system temperature and gain calibration was initially calculated using the EVN \textsc{parseltongue}~\citep{2005ADASS} pipeline written in \textsc{python} by Cormac Reynolds. The \textsc{aips} task \textsc{fring} was used to solve for the delay across the basebands with the fringe finder. Then, combining the basebands to give a better signal to noise, the phase, rates and delay of the phase calibrator were solved again using \textsc{fring}. A self-calibrated image of the phase calibrator was produced using the Caltech VLBI Software Package \textsc{difmap}~\citep{1997ASPC..125...77S}, enabling further calibration using \textsc{aips} to be performed.  The calibrated \textit{uv} data of GRS\,1915+105 was then Fourier transformed and the \textsc{clean} algorithm was applied using the \textsc{aips} task \textsc{imagr}. 

The radio image of GRS\,1915+105 on 2006 April 20~--~21 is shown in Fig.~\ref{grs1915} using a \textit{uv} weighting robustness parameter of 0~\citep{1995AAS...18711202B}. The source had a position of R.A.~$19^{\rm h}\,~15^{\rm m}\,~(11.548\pm0.001)^{\rm s}\,$ and Dec.~$10^{\circ}\,~56^{'}\,~(44.71\pm0.01)^{\farcs}$~(J2000). The position is consistent with that expected from the known proper motion~\citep{2005MNRAS.363..867M}.

The sources appears marginally resolved and was deconvolved from the beam using the \textsc{aips} task \textsc{jmfit} (the FWHM was $9.6\times6.5$~mas). This revealed an extended component estimated at $2.70\pm0.10\times1.2\pm0.05$~mas (FWHM), with a position angle of $140~(\pm2)^\circ$. This is similar to the P.A. of the large scale jets previously observed (e.g.~\cite{1999MNRAS.304..865F}). The total integrated radio flux density was $11.1~(\pm0.6)$~mJy. 

\subsection{Ryle Radio Telescope and RXTE monitoring of GRS\,1915+105}

The Ryle Radio Telescope and the RXTE all sky monitor\footnote{quick-look results provided by the ASM/RXTE team} regularly observes GRS\,1915+105. Fig.~\ref{monitor_grs1915+105} shows the XRB flux density between 2006 Jan - April at 15 GHz\ and 2 - 10 keV in the bottom and middle plots respectively. The top plot in Fig.~\ref{monitor_grs1915+105} shows the X-ray spectral hardness radio~$\left(\frac{5-12keV}{3-5keV}\right)$.

The date of the e-VLBI observation (MJD 53846), is marked. The flux entered a period of relative radio quietness in the two weeks before the e-VLBI observation. During the observation, the ASM count rate was about 40~s$^{-1}$, which is $\sim0.5$~crab~\citep{1996ApJ...469L..33L}. The X-ray spectral hardness changed just before the epoch of the observation to a slightly softer state.

\section{Discussion amd Conclusions}

\begin{figure}
\centering
\rotatebox{270}{\mbox{\resizebox{3cm}{!}{\includegraphics{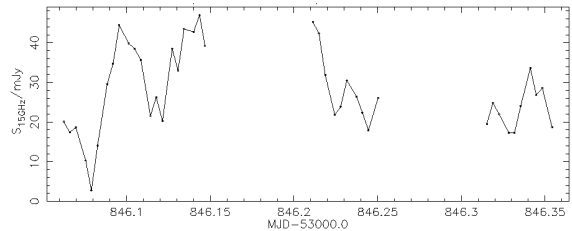}}}}
\caption{\label{21april_grs1915+105}Ryle Radio Telescope 15 GHz flux monitoring of GRS\,1915+105, observed at 2006 April 21 01:27 - 08:32 UT.}
\end{figure}

Fig.~\ref{21april_grs1915+105} shows the Ryle Radio Telescope data on 2006 April 21 between 01:27 - 08:32 UT. A flare of $\>40$~mJy was detected, which quickly decayed to $\sim20$~mJy within 4.5 hours. Assuming that the flare expands isotropically; the minimum energy in the magnetic field and energetic electrons can be calculated assuming equipartition within a synchrotron radiation field~\citep{2006pp381..419}. For a distance of 11 kpc~\citep{1999MNRAS.304..865F} the minimum energy is $2\times10^{41}$~ergs. Using the deconvolved size from the image rather than assuming spherical expansion, we find a minimum energy of $1\times10^{40}$~ergs, a lower value due to the source being collimated.
 
The radio emission for this and similar weak flares (see Fig.~\ref{monitor_grs1915+105}) decays rapidly ($<1$day). This is unlike the major flares studied by the VLA and MERLIN \citep{1999MNRAS.304..865F, 1994Natur.371...46M, 2005MNRAS.363..867M} where the decay is over several days and the ejecta can be followed for up to 2 months after the flare. The behaviour of the strong flare is consistent with the shock-in-jet model~\citep{2005MNRAS.363..867M}; however the short flares seem to show the charateristic of an expanding source without  continuous ejection of relativistic electrons. 

The relationship between the radio and X-ray flare is consistent with that for other black holes in the hard state~\citep{2003MNRAS.344...60G}. Such sources have compact jets and flat spectra. The spectrum measured in our observations between 5 and 15 GHz is flat or slightly inverted, and futhermore the source is aligned with the P.A. of major ejections observed by~\cite{1994Natur.371...46M}. This further supports the idea that radio jets are present when X-ray binaries are in the low/hard state. We note that though still in the hard state, figure~\ref{monitor_grs1915+105} shows that the hardness ratio falls slightly in coincidence with the occurrence of a weak radio flare.

The use of e-VLBI enabled us to obtain images within approximately a day of the VLBI run, rather than the many weeks needed for conventional recording based observations. It is possible to shorten the time between observations and image production even further, so that strategic decisions on future observations can be made for rapidly changing sources. The correlator needs to be stopped before correlated data can be off-loaded. Suitable gaps in the observing schedule could enable this to happen rather than waiting until the end of the run. The time to convert the data to an \textsc{aips} data file could be reduced by software improvements and finally avoidance of weekends would increase efficency.

This initial e-VLBI observation showed that the work load at observatories is decreased, while the load on correlator staff is increased considerably. Due note of this should be taken for resource allocation.

This work clearly shows the ability of the e-EVN to produce high resolution radio maps in real time, hence eliminating the need of tape\,/\,disc recording. In the future, e-VLBI transmission rates will keep increasing with network development, yielding higher sensitivities and longer baselines will be achieved with the addition of more telescopes to the network. Announcements of opportunity with information on applications are made on the e-VLBI web site (http://www.evlbi.org/evlbi) currently every $\sim$~2 months. This is a positive step in the development of a more dynamic and flexible network.

\section{Acknowledgements}

APR acknowledges support from a PPARC studentship during this research. The European VLBI Network is a joint facility of European, Chinese, South African and other radio astronomy institutes funded by their national research councils. e-VLBI developments in Europe are supported by the EC DG-INFSO funded Communication Network Development project,``EXPReS", Contract No. 02662 and the U.K. funded ESLEA project. The Ryle Radio Telescope is operated by the University of Cambridge and supported by PPARC. The X-ray data was provided by the ASM/RXTE teams at MIT and at the RXTE SOF and GOF at NASA's GSFC.

\bibliographystyle{mn2e}
\bibliography{File_not_for_review}

\label{lastpage}

\end{document}